\documentclass{article}

\usepackage{arxiv}

\usepackage[utf8]{inputenc} 
\usepackage[T1]{fontenc}    
\usepackage{hyperref}       
\usepackage{url}            
\usepackage{booktabs}       
\usepackage{amsfonts}       
\usepackage{nicefrac}       
\usepackage{microtype}      
\usepackage{lipsum}		
\usepackage{setspace}
\usepackage{graphicx}
\usepackage{doi}
\usepackage[numbers,sort&compress]{natbib}

\title{Kagome Topology in Two-Dimensional Noble-Metal Monolayers}

\author{%
\parbox{0.95\linewidth}{\centering
\href{https://orcid.org/0000-0002-9988-5202}{\includegraphics[scale=0.09]{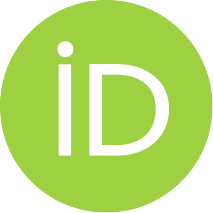}\hspace{1mm}}Carlos M.~O.~Bastos\textsuperscript{1,*},
\href{https://orcid.org/0009-0005-2176-678X}{\includegraphics[scale=0.09]{icons/orcid.pdf}\hspace{1mm}}Emanuel J.~A.~dos Santos\textsuperscript{2},
\href{https://orcid.org/0000-0002-8366-7227}{\includegraphics[scale=0.09]{icons/orcid.pdf}\hspace{1mm}}Jos\'e A.~dos S.~Laranjeira\textsuperscript{3},
\href{https://orcid.org/0000-0003-4699-5886}{\includegraphics[scale=0.09]{icons/orcid.pdf}\hspace{1mm}}Kleuton A.~L.~Lima\textsuperscript{4},
\href{https://orcid.org/0000-0001-5934-8528}{\includegraphics[scale=0.09]{icons/orcid.pdf}\hspace{1mm}}Alexandre C.~Dias\textsuperscript{1}, 
\href{https://orcid.org/0000-0003-0145-8358}{\includegraphics[scale=0.09]{icons/orcid.pdf}\hspace{1mm}}Douglas S.~Galv\~ao\textsuperscript{4},
and
\href{https://orcid.org/0000-0001-7468-2946}{\includegraphics[scale=0.09]{icons/orcid.pdf}\hspace{1mm}}Luiz A.~Ribeiro Jr\textsuperscript{2,$\dag$} \\
\vspace{0.6em}
{\normalfont\normalsize
\textsuperscript{1}Institute of Physics and International Center of Physics, University of Bras\'ilia, Bras\'ilia 70919-970, DF, Brazil\\
\textsuperscript{2}Computational Materials Laboratory, LCCMat, Institute of Physics, University of Bras\'ilia, 70910-900, Bras\'ilia, Federal District, Brazil\\
\textsuperscript{3}Modeling and Molecular Simulation Group, S\~ao Paulo State University (UNESP), School of Sciences, Bauru 17033-360, SP, Brazil\\
\textsuperscript{4}Department of Applied Physics and Center for Computational Engineering and Sciences, State University of Campinas, Campinas, 13083-859, SP, Brazil\\
\vspace{0.6em}
\href{https://scholar.google.com/citations?user=nkyVKy4AAAAJ\&hl=pt-PT}{\includegraphics[scale=0.05]{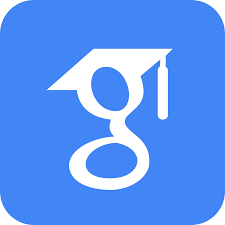}} \href{https://www.linkedin.com/in/carlos-maciel-de-oliveira-bastos-b6911432a}{\includegraphics[scale=0.05]{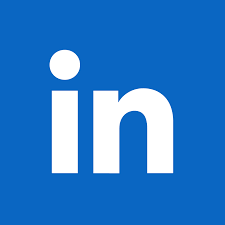}}\hspace{0.1cm}\texttt{\textsuperscript{*}carlos.bastos@unb.br} \\
\vspace{0.1cm}
\href{https://scholar.google.com/citations?user=EgsxcaUAAAAJ\&hl=pt-BR}{\includegraphics[scale=0.05]{icons/gscholar.png}} \href{www.linkedin.com/in/luiz-ribeiro-164221225}{\includegraphics[scale=0.05]{icons/linkedin.png}}\hspace{0.1cm}\texttt{\textsuperscript{$\dag$}ribeirojr@unb.br}\\
}
}%
}

\renewcommand{\shorttitle}{Carlos M.~O.~Bastos et al.}

\hypersetup{
pdftitle={kagome-goldene},
pdfsubject={kagome, goldene},
pdfauthor={Carlos M. O. Bastos, Luiz A. Ribeiro Jr},
pdfkeywords={Two-dimensional metals, Kagome lattice, Goldene, Lattice dynamics, Strain engineering, First-principles calculations.},
}

\begin{document}
\maketitle

\onehalfspacing

\begin{abstract}
Two-dimensional (2D) metallic lattices with kagome topology provide a unique platform for exploring the interplay between geometric frustration, reduced coordination, and lattice stability in elemental systems. Motivated by the recent experimental realization of atomically thin gold layers and kagome goldene, we present a first-principles investigation of free-standing kagome monolayers of Cu, Ag, and Au. Using density functional theory combined with lattice dynamics and ab initio molecular dynamics, we systematically assess their structural, mechanical, dynamical, and thermal stability. All kagome monolayers satisfy the 2D Born criteria and exhibit relatively low in-plane stiffness compared to graphene and hexagonal goldene, reflecting the porous nature of the kagome lattice and its metallic bonding. Among the three systems, the Au-based lattice displays the highest in-plane Young’s modulus. Phonon calculations reveal that the unstrained kagome phase is dynamically unstable for all metals. However, a moderate biaxial tensile strain of 5\% stabilizes the Ag and Au monolayers, while Cu retains residual unstable modes. Finite-temperature simulations further show that Cu rapidly reconstructs toward a trigonal lattice, Ag remains metastable at low temperature but collapses at room temperature, and Au exhibits competing kagome and trigonal motifs at 300 K, indicating near-degeneracy between these phases. These results establish that strain engineering and atomic size are key determinants of the stability of metallic kagome monolayers and provide guidance for future substrate-supported realizations.
\end{abstract}


\section{Introduction}

The isolation of atomically thin metals has recently expanded the landscape of two-dimensional (2D) materials beyond covalent and van der Waals crystals \cite{kashiwaya2024synthesis,sharma2022synthesis,ramachandran2024gold,preobrajenski2024boron,mironov2024graphene}. Among elemental systems, gold has emerged as a particularly robust platform \cite{ramachandran2024gold}. Recent experimental breakthroughs have demonstrated the realization of single-atom-thick gold layers, commonly referred to as goldene, obtained through selective etching and confinement strategies that stabilize a close-packed triangular monolayer \cite{kashiwaya2024synthesis,sharma2022synthesis}. These studies established that goldene preserves metallic character while exhibiting enhanced mechanical rigidity and stability relative to its bulk counterpart, thereby opening a route toward truly 2D noble-metal systems. Importantly, theoretical studies have also demonstrated the stability and enhanced mechanical properties of monolayer goldene \cite{mortazavi2024goldene,pereira2025does,dos2025exploring,abidi2025electronic,abidi2025atomically,abidi2024gentle}.

Beyond dense triangular lattices, reduced-coordination geometries provide access to qualitatively different physical regimes \cite{lubensky2015phonons}. In this context, kagome lattices occupy a central position due to their intrinsic geometric frustration, which gives rise to flat electronic bands, Dirac features, and soft lattice modes \cite{zhang20252d,jing2018two,ghimire2020topology}. While kagome physics has been extensively explored in compounds and artificial lattices \cite{guo2009topological,jo2012ultracold,wang2024topological}, its realization in elemental 2D metals is still in its early stages \cite{preobrajenski2024boron,tian2025kagome}. Very recently, kagome goldene was experimentally achieved via line-graph epitaxy, revealing flat bands and Dirac nodal features in a purely metallic kagome network \cite{tian2025kagome}. This advance highlights the feasibility of engineering topologically nontrivial lattices in atomically thin metals, while simultaneously raising fundamental questions regarding their intrinsic stability.

From a lattice-dynamical perspective, kagome geometries are inherently susceptible to soft modes \cite{chen2025absence}. This stems from their reduced coordination and frustrated bonding. Recent theoretical work has shown that even tensile strain can profoundly modify the phonon spectrum of atomically thin (trigonal) metallenes \cite{abidi2024gentle}. Slight tensile strain suppresses imaginary modes and stabilizes otherwise unstable configurations. This "gentle-tension" mechanism provides a physically transparent method to stabilize porous metallic 2D lattices without altering their topology or chemistry.

Motivated by recent developments on the synthesis of gold-based kagome lattices (kagome-goldene) \cite{tian2025kagome}, we present a first-principles study of free-standing kagome monolayers composed of Cu, Ag, and Au. We use density functional theory (DFT), phonon calculations, and ab initio molecular dynamics (AIMD) to assess the mechanical response and stability of these materials under strain and at finite temperature. Our results clarify the effects of atomic size and relativistic contributions on the stabilization of kagome metallic lattices. We also provide a stability framework relevant to future substrate-supported and strain-engineered realizations.

\section{Methodology}

All calculations were performed within the framework of DFT~\cite{hohenbergB8641964}, as implemented in the Vienna Ab Initio Simulation Package (VASP)~\cite{kresse1996efficient,kresse1996efficiency}, version 6.5.0. The Kohn–Sham equations~\cite{kohnA11331965} were solved using the semilocal exchange–correlation functional proposed by Perdew, Burke, and Ernzerhof (PBE)~\cite{perdew38651996c}. The interaction between core and valence electrons was described using the projector augmented-wave (PAW) method~\cite{blochl179531994,kresse1999ultrasoft}. Spin–orbit coupling (SOC) effects were included using the second-variational approach~\cite{koelling31071977} as implemented in VASP~\cite{steiner2244252016}. SOC was accounted for in all calculations, except for phonon and AIMD simulations, where it was neglected due to the associated computational cost.

Structural optimizations minimized the total energy with respect to both atomic positions and in-plane lattice parameters, proceeding until the residual Hellmann–Feynman forces on each atom were below 0.01 eV/\r{A}. The plane-wave basis-set cutoff energy was set to twice the maximum recommended value (ENMAX) from the respective POTCAR files, resulting in cutoff energies of 600 eV, 500 eV, and 460 eV for Cu, Ag, and Au. Brillouin-zone integrations used a $\Gamma$-centered Monkhorst–Pack grid~\cite{monkhorst51881976} of $9 \times 9 \times 1$ for all systems.

Elastic stiffness constants were calculated from the optimized structures using the finite-difference method in VASP~\cite{lepage1041042002}, where slight strains are applied to the lattice and the resulting stress–strain response is measured. Phonon dispersion relations were found using density functional perturbation theory (DFPT)~\cite{gajdos0451122006}, where the dynamical matrix comes from the second derivatives of the total energy with respect to atomic positions. For both elastic and phonon calculations, the same exchange–correlation functional, cutoff energy, and k-point sampling as in the structural optimizations were used.

Finite-temperature stability was investigated through ab initio molecular dynamics simulations performed on a $4 \times 4 \times 1$ supercell within the canonical (NVT) ensemble, using a Nosé–Hoover thermostat~\cite{nose5111984} as implemented in VASP. Simulations were carried out at temperatures of 50 K and 300 K. The plane-wave cutoff energy was kept identical to that used in the static calculations. At the same time, Brillouin-zone sampling was restricted to the $\Gamma$ point.

\section{Results}

\begin{figure}[!htb]
    \centering
    \includegraphics[width=\linewidth]{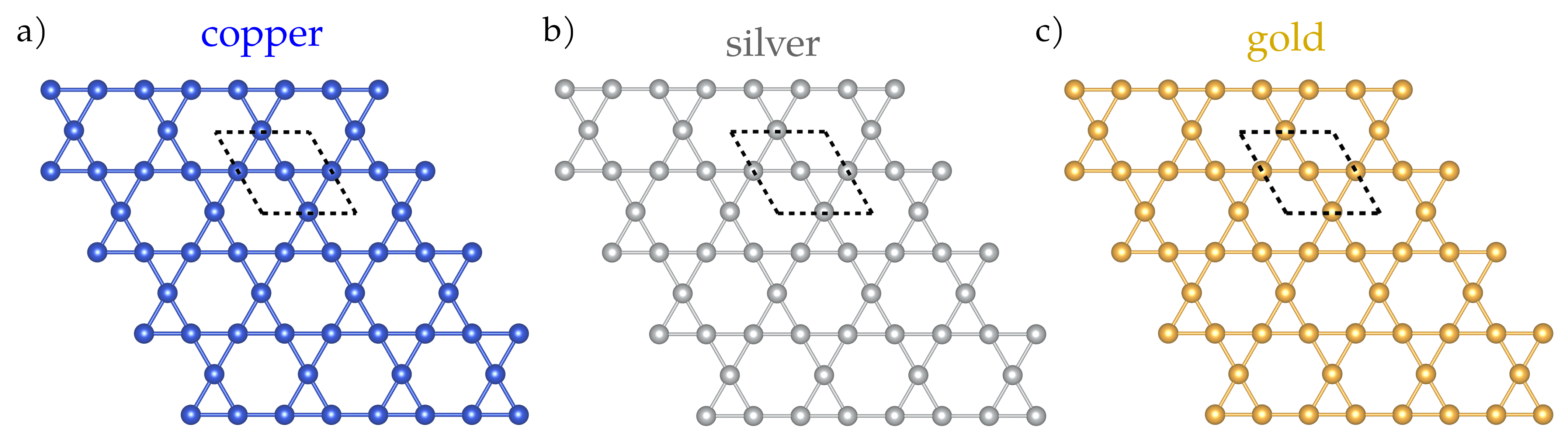}
   \caption{Optimized atomic structures of free-standing kagome monolayers composed of (a) Cu, (b) Ag, and (c) Au. Dashed hexagons indicate the primitive unit cell containing three atoms. The kagome topology is formed by corner-sharing triangular units, resulting in a coordination number of four. All structures are shown after full in-plane relaxation, with a fixed vacuum spacing along the out-of-plane direction.}
\label{fig:structures}
\end{figure}

We begin by presenting the structural properties. Specifically, the free-standing kagome monolayers of Cu, Ag, and Au crystallize in the hexagonal space group $P6/mmm$. Each primitive unit cell contains three atoms, with a coordination number of four (see Figure \ref{fig:structures}). This reduced coordination, intrinsic to the kagome topology, distinguishes these systems from close-packed triangular monolayers such as goldene \cite{mortazavi2024goldene,pereira2025does}, and it also plays a central role in their mechanical and dynamical behavior. By symmetry, the lattice parameters satisfy $a=b$, while the interaxial angles are $\alpha=\beta=90^\circ$ and $\gamma=120^\circ$.

Figure~\ref{fig:structures} shows the relaxed atomic structures of the kagome monolayers for Cu, Ag, and Au, highlighting the characteristic corner-sharing triangular motifs and the corresponding primitive unit cell. While the overall kagome geometry is preserved across all three systems, differences in lattice constants and bond strengths foreshadow distinct trends in mechanical response and stability, as discussed in the following sections.

To model the 2D nature of the systems within periodic boundary conditions, a vacuum spacing of 20\r{A} was introduced along the out-of-plane direction. The lattice parameter $c$ was kept fixed during structural optimization to suppress spurious interactions between periodic images, while all in-plane lattice parameters and atomic positions were fully relaxed. The resulting equilibrium lattice constants $a$ are summarized in Table~\ref{table:lattice}.

\begin{table}[!htb]
\centering
\caption{Structural and in-plane elastic properties of free-standing metallic kagome monolayers. The equilibrium lattice constant $a$ and the 2D Young's modulus $E$ are reported together with the 2D elastic stiffness constants $C_{11}$ and $C_{12}$ (all in N/m). Importantly, these values indicates that the 2D Born stability conditions are satisfied (i.e., $C_{11}>0$ and $C_{11}-C_{12}>0$) for all systems.}
\label{table:lattice}
\begin{tabular}{lcccc}
\toprule
& \textbf{$a$ (\AA)} & \textbf{$E$ (N/m)} & \textbf{$C_{11}$ (N/m)} & \textbf{$C_{12}$ (N/m)} \\
\midrule
Cu & 4.74 & 31.6 & 42.3 & 21.2  \\
Ag & 5.45 & 27.8 & 33.1 & 13.2  \\
Au & 5.30 & 55.4 & 65.9 & 26.3 \\
\bottomrule
\end{tabular}
\end{table}

The optimized lattice parameters follow the expected trend, primarily driven by atomic size, with Cu exhibiting the smallest in-plane lattice constant, followed by Au and Ag. The lattice parameter of Au ($a=5.30$ \r{A}) is smaller than that of Ag ($a=5.45$ \r{A}), despite the larger atomic number of gold. This behavior is well established in gold-based systems \cite{lee2003geometrical}. It can be attributed to relativistic effects, in particular the contraction of the $6s$ orbital and the enhanced $5d$--$6s$ hybridization, which strengthen Au--Au bonding and lead to shorter equilibrium bond lengths \cite{schwerdtfeger2002relativistic}. In contrast, relativistic stabilization is significantly weaker in silver, leading to a larger equilibrium lattice parameter.

To assess the mechanical response and stability of the metallic kagome monolayers, we computed the independent in-plane elastic stiffness constants $C_{11}$ and $C_{12}$. Owing to the hexagonal symmetry of the kagome lattice, the in-plane shear modulus $C_{66}$ is not independent and is given by the relation
\begin{equation}
C_{66} = \frac{C_{11}-C_{12}}{2}.
\end{equation}

Furthermore, for 2D systems with hexagonal symmetry, the in-plane Young’s modulus $E_{2D}$ can be expressed in terms of the elastic constants as
\begin{equation}
E_{2D} = \frac{C_{11}^2 - C_{12}^2}{C_{11}}.
\end{equation}

Using these relations, we obtained in-plane Young’s moduli of 31.6 N/m, 27.8 N/m, and 55.4 N/m for the kagome monolayers of Cu, Ag, and Au, respectively. The full set of elastic stiffness constants and derived mechanical properties is summarized in Table~\ref{table:lattice}. Among the three systems, the Au monolayer exhibits a significantly larger Young’s modulus, indicating a stiffer in-plane mechanical response compared to its Cu and Ag counterparts. This enhanced stiffness can be attributed to stronger interatomic bonding in gold, which is widely associated with relativistic effects, particularly the stabilization of the $5d$--$6s$ hybridization that strengthens Au--Au interactions.

Despite this relative stiffness, the calculated in-plane Young’s moduli of the kagome monolayers (27--55 N/m) are approximately one order of magnitude smaller than that of graphene ($E \approx 340$ N/m)~\cite{lee3852008}. This pronounced difference originates from two main factors. First, graphene is stabilized by strong, highly directional $sp^2$ covalent bonds, which provide exceptional in-plane rigidity, whereas bonding in the kagome metal lattices is predominantly metallic and less directional. Second, the kagome topology is intrinsically porous, consisting of corner-sharing triangular units that allow for rotational and distortive degrees of freedom under strain. This open framework reduces the effective resistance to deformation compared to the dense honeycomb lattice of graphene.

A more direct and physically meaningful comparison can be made between the kagome-goldene monolayer investigated here and the recently reported triangular goldene phase. Previous DFT studies have shown that goldene exhibits a substantially larger in-plane Young’s modulus, on the order of $\sim$100--200 N/m, reflecting its close-packed triangular geometry and higher coordination number~\cite{mortazavi2024goldene}. In contrast, the kagome-goldene monolayer displays a significantly reduced Young’s modulus of 55.4 N/m, highlighting the strong influence of lattice topology on the mechanical response. While both systems are composed of the same chemical element and benefit from relativistic stabilization of Au--Au bonding, the reduced coordination number and porous nature of the kagome lattice introduce additional low-energy deformation pathways that markedly soften the in-plane stiffness. 

Mechanical stability was further evaluated using the Born stability criteria for two-dimensional hexagonal lattices, which require (i) $C_{11}>0$ and (ii) $C_{11} > |C_{12}|$~\cite{malyi248762019}. All three kagome monolayers satisfy these conditions, confirming that, despite their reduced stiffness and open topology, the metallic kagome lattices are mechanically stable in the harmonic regime. These results establish a mechanically viable foundation for the subsequent analysis of lattice dynamics and finite-temperature stability.

\begin{figure}[!htb]
    \centering
    \includegraphics[width=\linewidth]{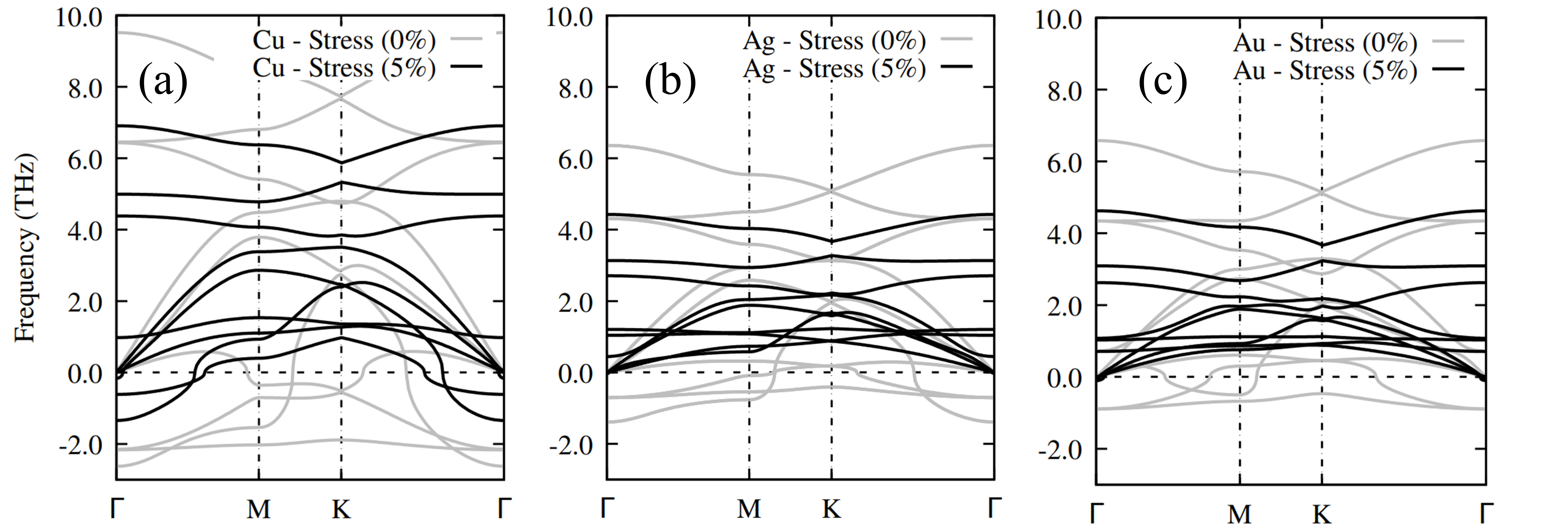}
   \caption{Phonon dispersion relations of free-standing kagome monolayers composed of (a) Cu, (b) Ag, and (c) Au along the high-symmetry path $\Gamma$--M--K--$\Gamma$. Gray curves correspond to the unstrained lattices, while black curves show the phonon spectra under 5\% biaxial tensile strain.}
\label{fig:phonons}
\end{figure}

The dynamical stability of the free-standing kagome monolayers of Cu, Ag, and Au was investigated through their phonon dispersion relations, shown in Figure~\ref{fig:phonons}. In the absence of external strain, all three systems exhibit imaginary phonon frequencies, represented as negative values in the dispersion curves (grey), signaling dynamical instability of the unstrained kagome lattice. These unstable modes are predominantly located in the low-frequency region, reflecting the lattice's intrinsic softness, associated with reduced coordination and geometric frustration in the kagome topology.

It has been demonstrated that 2D metallic systems obtained from DFT often require external stabilization mechanisms to suppress soft modes and achieve dynamical stability. In particular, it was shown that the application of "gentle tensile strain" can effectively stabilize otherwise unstable 2D metallic lattices by modifying the curvature of the potential-energy surface and lifting low-energy vibrational instabilities \cite{abidi2024gentle}. Motivated by this insight, we applied a 5\% biaxial tensile strain to all kagome monolayers and recalculated their phonon dispersions.

Under biaxial strain, the phonon spectra of the Ag and Au kagome monolayers exhibit exclusively positive frequencies throughout the Brillouin zone, indicating that the applied strain is sufficient to fully suppress the imaginary modes and render these systems dynamically stable. In contrast, although the magnitude of the imaginary frequencies in the Cu kagome monolayer is significantly reduced under strain, residual unstable modes persist, indicating that strain alone is insufficient to stabilize the copper lattice.

The distinct behavior of Cu can be attributed to a combination of factors. Copper has a smaller atomic radius than silver and gold, which limits its ability to accommodate the open, porous kagome geometry. Moreover, the kagome lattice has a reduced coordination number of four, in contrast to close-packed triangular monolayers such as goldene, where each atom is sixfold coordinated~\cite{dos2025exploring,pereira2025does,mortazavi2024goldene}. This reduced coordination enhances lattice flexibility and amplifies soft vibrational modes, particularly for lighter elements with weaker relativistic bonding effects. As a result, while strain engineering is effective in stabilizing the kagome phases of Ag and Au, the Cu kagome monolayer remains dynamically unstable in the free-standing form.

\begin{figure}[!htb]
    \centering
    \includegraphics[width=\linewidth]{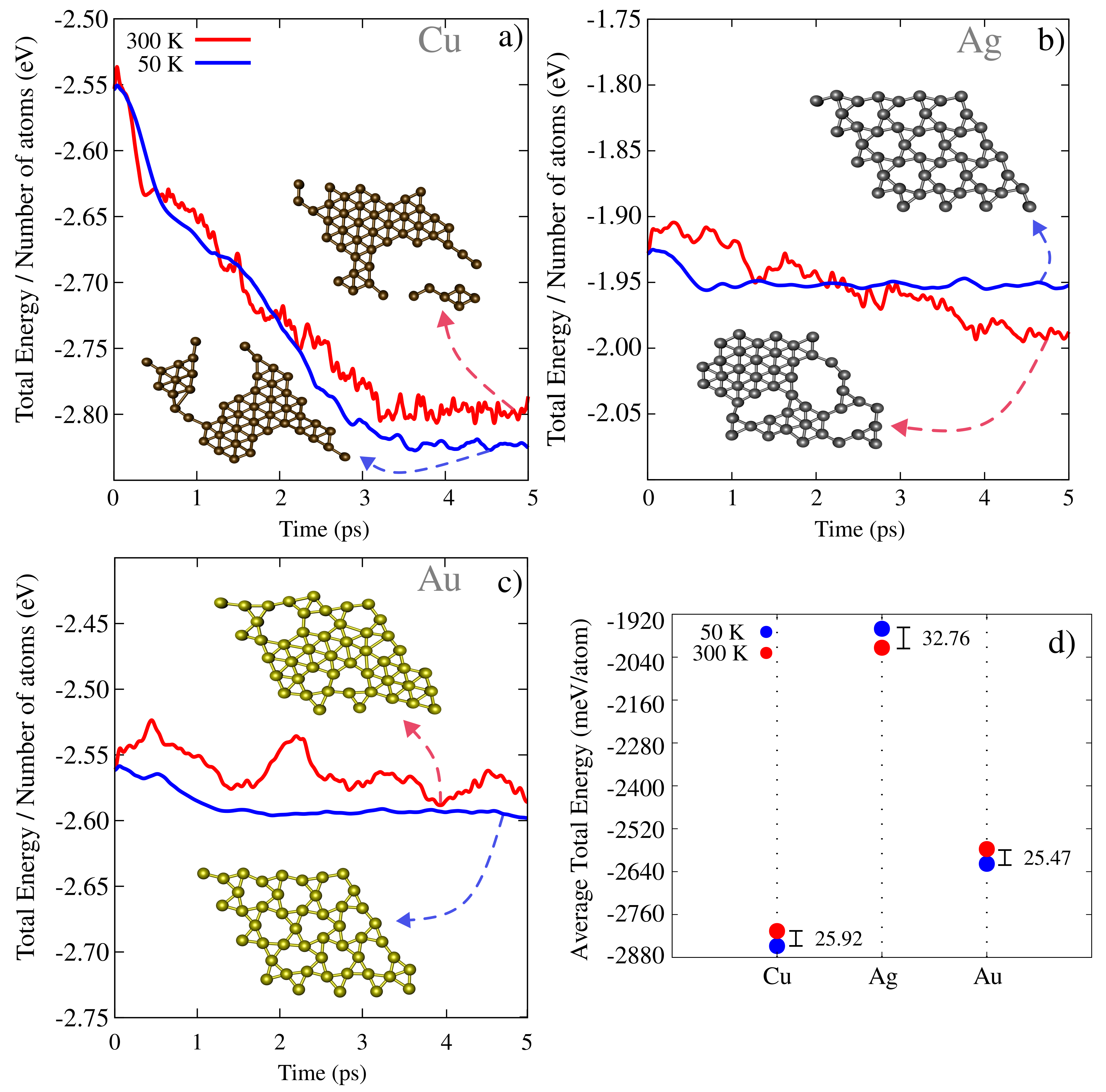}
 \caption{AIMD results for free-standing metallic kagome monolayers. Time evolution of the total energy per atom at 50 K (blue curves) and 300 K (red curves) for (a) Cu, (b) Ag, and (c) Au.}
\label{fig:aimd}
\end{figure}

Finally, to assess the thermal stability of the metallic kagome monolayers and the effect of temperature on their lattice dynamics, we performed AIMD simulations within the canonical (NVT) ensemble at 50 K and 300 K, as shown in Figure~\ref{fig:aimd}. The time evolution of the total energy, together with representative atomic configurations, provides direct insight into the metastability and reconstruction pathways of the kagome lattices.

For the Cu kagome monolayer, the kagome phase is unstable at both temperatures, as illustrated in Figure~\ref{fig:aimd}(a). In this case, the energy barrier associated with the open kagome topology is insufficient to preserve the lattice, leading to a spontaneous first-order structural transformation into a trigonal phase within approximately 3.5 ps. This rapid collapse indicates that thermal fluctuations, even at low temperatures, readily activate reconstruction pathways in copper due to its smaller atomic size and greater susceptibility to lattice distortions.

In contrast, the kagome monolayers of Ag and Au, which possess larger atomic radii compared to Cu, display enhanced structural robustness. At 50 K, thermal fluctuations induce local distortions but do not trigger a phase transition, thereby preserving the kagome topology in both metals. At 300 K, however, their behavior diverges markedly. For Ag, the increased thermal energy is sufficient to overcome the reconstruction barrier, resulting in a collapse of the kagome lattice into a trigonal structure, as shown in Figure~\ref{fig:aimd}(b). 

The Au kagome monolayer exhibits a more complex response, as shown in Figure~\ref{fig:aimd}(c). Although no complete structural transition is observed within the simulation time, significant lattice distortions develop at 300 K. The atomic configurations reveal the coexistence of kagome-like and trigonal-like motifs, indicating a competition between these two structural phases. This behavior suggests that the kagome and trigonal configurations of gold are energetically close, leading to a shallow energy landscape in which both motifs can be locally stabilized.

This interpretation is supported by the analysis of the average total energy over the dynamically stable portions of the trajectories, shown in Figure~\ref{fig:aimd}(d). The averaging windows were chosen as 3.5 ps, 4.0 ps, and 3.0 ps for Cu, Ag, and Au, respectively. For Ag, an apparent decrease in the average total energy is observed at 300 K relative to 50 K, confirming that thermal activation drives the system from a metastable kagome configuration toward a lower-energy trigonal phase. In contrast, the average energies of the Cu and Au monolayers at 300 K remain comparable to those at 50 K, apart from the expected increase due to thermal fluctuations. While Cu adopts the same trigonal phase at both temperatures, Au exhibits distinct structural states, retaining a kagome lattice at 50 K and evolving toward trigonal motifs at 300 K.

\section{Conclusion}

In summary, we performed a comprehensive first-principles investigation of free-standing metallic kagome monolayers composed of Cu, Ag, and Au, focusing on their structural, mechanical, and dynamical properties. Our results show that all three systems adopt a hexagonal kagome lattice with reduced coordination, resulting in relatively low in-plane stiffness compared with dense 2D metals and graphene. Among them, the kagome Au monolayer exhibits the largest Young’s modulus, reflecting stronger interatomic bonding driven by relativistic effects. Elastic stability analysis confirms that all systems satisfy the two-dimensional Born criteria, establishing a mechanically stable foundation despite the intrinsic softness of the kagome topology.

Phonon and finite-temperature analyses, however, reveal that mechanical stability alone is insufficient to guarantee dynamical and thermal stability in these systems. Unstrained kagome monolayers display soft phonon modes, which can be fully suppressed in Ag and Au by the application of moderate biaxial tensile strain. In contrast, residual instabilities persist in Cu. AIMD simulations further demonstrate that the kagome phase of Cu is unstable even at low temperature, while Ag and Au exhibit metastable kagome configurations whose survival depends sensitively on temperature. In particular, gold displays a unique regime of structural competition in which kagome and trigonal motifs coexist, highlighting the decisive role of lattice topology, atomic size, and relativistic bonding in stabilizing two-dimensional metallic kagome lattices. These findings provide a fundamental framework for the strain- and substrate-assisted realization of 2D kagome metals.

\section*{Acknowledgments}
This work was supported by the Brazilian funding agencies Fundaç\~ao de Amparo à Pesquisa do Estado de S\~ao Paulo - FAPESP (grant no. $2022/16509-9$). L.A.R.J. acknowledges the financial support from FAP-DF grants $00193.00001808$ $/2022-71$ and $00193-00001857/2023-95$, FAPDF-PRONEM grant $00193.00001247/2021-20$, and CNPq grants $350176/2022-1$ and $167745/2023-9$. A.C.D. acknowledges the financial support from FAP-DF grants $00193-00001817/2023-43$ and $00193-00002073/2023-84$, and CNPq grants $305174/2023-1$, $444069/2024-0$, and $444431/2024-1$. C.M.O.B, A.C.D. and L.A.R.J. acknowledge PDPG-FAPDF-CAPES Centro-Oeste grant $00193-00000867/2024-94$. The authors also express their gratitude to “Centro Nacional de Processamento de Alto Desempenho em S\~ao Paulo”
(CENAPAD-SP) and to the Lobo Carneiro HPC (NACAD) supercomputer at the Federal University of Rio de Janeiro (UFRJ).

\bibliographystyle{unsrtnat}
\bibliography{references}  

\end{document}